\documentclass[twocolumn,aps,floatfix,superscriptaddress]{revtex4}
\usepackage{amsmath,amssymb,eucal,graphicx}

\begin{document}
\title{Molecular Spiders in One Dimension}

\author{Tibor Antal}\affiliation{Program for Evolutionary Dynamics,
  Harvard University, Cambridge, MA 02138, USA}
  \author{P. L. Krapivsky}\affiliation{Department of Physics and Center for Molecular
Cybernetics, Boston University, Boston, MA 02215, USA}
 \author{Kirone Mallick}\affiliation{Service de Physique Th\'eorique, Cea Saclay,
  91191 Gif, France}

\begin{abstract}
Molecular spiders are synthetic bio-molecular systems which have ``legs'' made of  short single-stranded segments of DNA. Spiders move on a surface covered with  single-stranded DNA segments complementary to legs. Different mappings are established between various models of spiders and simple exclusion processes. For spiders with simple gait  and varying number of legs we compute the diffusion coefficient; when the hopping is biased we also compute their velocity.
\end{abstract}

\maketitle

\section{Introduction}

In recent years, chemists have constructed a number of synthetic
molecular systems which can move on surfaces and tracks (see e.g.  \cite{seeman,pierce,DNA_motor} and a comprehensive review \cite{review}). One such object
is a multi-pedal molecular spider whose legs are short single-stranded
segments of DNA \cite{exp}. These spiders can move on a surface
covered with single-stranded DNA segments, called substrates. The
substrate DNA is complementary to the leg DNA. The motion proceeds as
legs bind to the surface DNA through the Watson-Crick mechanism, then
dissociate, then rebind again, etc.  More precisely, a bond on the substrate with an attached leg is first cleaved \cite{exp}, and the leg then dissociates from the affected substrate (which we shall call product). The leg then rebinds again to the new
substrate or to the product leading to the motion of the spider.

The rate of attachment of a leg of the spider to the substrate and the
rate of detachment from the substrate are different from the
corresponding rates involving the product instead of the
substrate. Hence for the proper description of the motion of a single
spider one must keep track of its entire trajectory. This memory
requirement makes the problem non-Markovian \cite{markov,van} and
generally intractable analytically even in the case of a single spider. (We shall address this problem in a separate paper \cite{memospiders}.)
Many interacting spiders add another level of complexity. Even
if the rates were the same for the substrate and the product, the
properties of the spider (e.g., the nature of its gait) represent
another major challenge. Here we separate this latter issue from the
rest: we investigate how the structure of the spider affects its
characteristics (velocity and diffusion coefficient).  Further, we
consider spiders with idealized gait --- the goal is not to mimic
complicated (and poorly known) gait of molecular spiders but to show
that spiders' macroscopic characteristics are very sensitive to their
structure and gait.

We shall mostly focus on a single spider moving on a lattice. We shall also
assume that the rate of attachment greatly exceeds the rate of
detachment. In this situation the relative time when one leg is
detached is negligible and hence the possibility that two or more legs
are detached simultaneously can be disregarded. Therefore the present
model posits that spiders remain fully attached and never leave the
surface \cite{models}.  (This differs from the actual situation when a
few legs may be simultaneously detached.)

The spiders are defined as follows. Legs can jump independently at constant rates if they do not violate the restrictions below. We mainly consider symmetric spiders where we set all these rates to one, or biased spiders whose legs can only move to the right at rate one, but some special gaits are also investigated. The fundamental restriction on the spider's motion is the exclusion principle: Two legs cannot bind to the same site. 
Additional constraints keep the legs close to each other. We mainly consider two types of spiders with the simplest feasible constraints:

{\bf Centipedes} (or local spiders). A leg of a centipede can step to nearest neighbor sites provided that it remains within distance $s$ from the {\em adjacent} legs. (This threshold is assumed to be the same for each pair of adjacent legs.)

{\bf Spiders} (or global spiders). Legs of these spiders can step to nearest neighbor sites as long as {\em all} legs remain within distance $S$.

The above properties of the gait guarantee that in {\em one dimension} the order
of the legs never changes. The above constraints seem equally natural in one
dimension, while in two dimensions the global constraint appears more
reasonable.

We shall also briefly discuss a third type of spider where the nearest neighbor restriction on the hopping is relaxed. For these {\em quick spiders}, legs can step {\em anywhere} within distance $S$ from {\em all} legs. Quick spiders have been proposed and studied numerically in Ref.~\cite{OS}. 

The above assumptions about the gait and the disregard of memory effects leave little hope for quantitative modeling, but simplicity can help to shed light on qualitative behaviors. Therefore we study in depth a single spider with aforementioned gait moving on a one-dimensional lattice, and more briefly probe the influence of the gait and many-spider effects.

The rest of this paper is organized as follows. In Sec.~\ref{bi}, we analyze 
bipedal spiders (i.e., spiders with two legs). This framework provides a useful laboratory to probe various techniques. Bipedal spiders also closely resemble molecular motors \cite{motors} and the methods developed for studying molecular motors are  fruitful for studying individual spiders \cite{FK,D}.  In Sec.~\ref{multi} we examine multi-pedal spiders (i.e.,  spiders  with $L\ge 3$ legs).   We show that the spider with local constraint and $s=2$  is isomorphic to a simple exclusion process (SEP) on a line with $L-1$ sites and open boundary conditions; an even simpler isomorphism exists between spiders with global constraint and the SEP on the ring. These connections allow us to extract some spider characteristics from results about the SEP. Quick spiders are briefly investigated in Sec.~\ref{long}. In Sec.~\ref{inter} we show that the behavior of many interacting spiders can also be understood, at least in the practically important situation of low spider density, via the connection with the SEP. Finally we stress limitations of our analysis and discuss possible extensions in Sec.~\ref{disc}.

\section{Bipedal Spider}
\label{bi}

For bipedal spiders, the local and global constraints are equivalent,
$s\equiv S$. For the simplest mobile bipedal spider, the allowed
distance between the legs is one or two lattice spacing,
i.e.,  $s=2$. Two  possible configurations are (up to translation)
\begin{equation}
\label{2} 
\ldots\circ\bullet\bullet\circ\ldots \qquad {\rm and}\qquad
\ldots\circ\bullet\circ\bullet\circ\ldots
\end{equation}
where we denote empty sites by `$\circ$' and filled sites (to which
the legs are attached) by `$\bullet$'. There are obvious back and
forth transitions between these configurations:
\begin{equation*}
\bullet\bullet\circ\Longleftrightarrow\bullet\circ\bullet\qquad {\rm
and}\qquad \circ\bullet\, \bullet\Longleftrightarrow\bullet\circ\bullet
\end{equation*}
For symmetric spiders each leg jumps at rate one when possible, hence all the above four elementary moves happen at rate unity. The diffusion coefficient of this bipedal spider is
\begin{equation}
\label{D2} 
D_2=\frac{1}{4}
\end{equation}
To put this in perspective, we note that the diffusion coefficient of
a random walker which hops to the right and left with unit rates is
$D=1$. Thus adding a leg and requiring the legs to stay within distance two to each other reduces the diffusion coefficient by a factor 4.

Generally for symmetric bipedal spiders with arbitrary $s$, there are $s$ possible configurations
$\mathcal{C}_\ell$ labeled by the inter-leg distance, $\ell=1,\ldots,s$. The transitions are
\begin{equation*}
\mathcal{C}_1\Longleftrightarrow \mathcal{C}_2\Longleftrightarrow
\ldots\Longleftrightarrow\mathcal{C}_s
\end{equation*}
The diffusion coefficient $D_s$ of this bipedal spider is
\begin{equation}
\label{Ds} 
D_s=\frac{1}{2}\left(1-\frac{1}{s}\right)
\end{equation}

The above results apply to symmetric bipedal spiders which hop to the left and
right with equal rates. Molecular motors usually undergo directed
motion \cite{motors}, and one of the goals of future research is
to control spiders to move preferentially in a certain
direction. Here we analyze such directed motion theoretically. For
concreteness, we focus on the extreme bias when each leg can only hop to right at rate one.
For instance, for the bipedal spider the most compact
configuration evolves via
$\bullet\bullet\circ\Longrightarrow\bullet\circ\bullet$\,; the process
$\circ\bullet\bullet\Longrightarrow\bullet\circ\bullet$ involves
hopping to the left and therefore it is forbidden in the biased case.

For biased bipedal spiders the velocity and the diffusion coefficient
are given by
\begin{equation}
\label{VDL} 
V_s=1-\frac{1}{s}\,\,,\qquad
D_s=\frac{1}{3}\left(1-\frac{1}{s}\right)\left(1-\frac{1}{2s}\right)
\end{equation}

In this section we give a pedestrian derivation of \eqref{Ds}. The
expressions \eqref{VDL} for velocity and diffusion coefficient can be
derived by utilizing the same technique; instead, we shall extract
them from more general results for lame spiders
(Sec.~\ref{lame}).

To set the notation and to explain how we compute the diffusion
coefficient we begin with a random walk (which is a one-leg
spider). Let $P_n(t)$ be the probability that the random walker is at
site $n$ at time $t$. This quantity evolves according to
\begin{equation}
\label{RW-master} 
\frac{d P_n}{dt}=P_{n-1}+P_{n+1}-2P_n
\end{equation}
One can solve this equation and then use that solution to extract the
diffusion coefficient. In the case of the spiders, however, master
equations generalizing (\ref{RW-master}) are much less tractable, and
therefore a more direct way of computing the diffusion coefficient is
preferable.  Here we describe one such approach \cite{van}. It
involves two steps.  First, one ought to determine the mean-square
displacement
\begin{equation}
\label{mean-def} 
\langle x^2\rangle=\sum_{n=-\infty}^\infty n^2 P_n
\end{equation}
Then the basic formula \cite{van}
\begin{equation}
\label{D-extract} 
D=\lim_{t\to \infty}\frac{\langle x^2\rangle}{2t}
\end{equation}
allows to extract the diffusion coefficient.

For the random walk, the mean-square displacement evolves according to
\begin{equation}
\label{mean-eq} 
\frac{d}{dt}\,\langle x^2\rangle =\sum_{n=-\infty}^\infty
n^2\left(P_{n-1}+P_{n+1}-2P_n\right)
\end{equation}
 Transforming  the first two sums we obtain
\begin{equation}
\label{-+} 
\sum_{n=-\infty}^\infty n^2 P_{n\mp 1}=\sum_{n=-\infty}^\infty (n\pm
1)^2 P_n
\end{equation}
These identities allow us to recast (\ref{mean-eq}) into
\begin{eqnarray*}
\frac{d}{dt}\,\langle x^2\rangle &=&\sum_{n=-\infty}^\infty
\left[(n+1)^2+(n-1)^2-2n^2\right]P_n\\ &=& 2\sum_{n=-\infty}^\infty
P_n=2
\end{eqnarray*}
where the last equality follows from normalization.  Thus $\langle
x^2\rangle=2t$. Plugging this into (\ref{D-extract}) we recover the
diffusion coefficient of the random walker $D=1$.

We now turn to the bipedal spider. We shall examine in detail only
symmetric hopping.

\subsection{Bipedal Spider with $s=2$}

For the bipedal spider with $s=2$ there are two possible spider
configurations.  Denote by $P_n(t)$ and $Q_n(t)$ the probabilities
that at time $t$ the spider is in respective configurations
\eqref{2}, namely
\begin{equation}
\label{PQ-def}
P_n={\rm Prob}[\bullet\, \bullet], \quad Q_n={\rm
Prob}[\bullet\circ\bullet],
\end{equation}
with the left leg being at site $n$. The governing equations for these
probabilities are
\begin{subequations}
\begin{align}
&\frac{d P_n}{dt}=Q_{n}+Q_{n-1}-2P_n
\label{Pn}\\
&\frac{d Q_n}{dt}=P_{n+1}+P_{n}-2Q_n
\label{Qn}
\end{align}
\end{subequations}
The mean position of the legs or the `center of mass' of the spider in a configuration corresponding to
$P_n$ (resp. $Q_n$) is located at $n+\frac{1}{2}$ (resp. $n+1$). Thus
the mean-square displacement is
\begin{equation}
\label{mean-2} 
\langle x^2\rangle=\sum_{n=-\infty}^\infty
\left[\left(n+\frac{1}{2}\right)^2 P_n+(n+1)^2Q_n\right]
\end{equation}
and it evolves according to
\begin{eqnarray*}
\frac{d}{dt}\,\langle x^2\rangle &=&\sum_{n=-\infty}^\infty
\left(n+\frac{1}{2}\right)^2(Q_{n}+Q_{n-1}-2P_n)\\
&&+\sum_{n=-\infty}^\infty (n+1)^2(P_{n+1}+P_{n}-2Q_n)
\end{eqnarray*}
Utilizing the same tricks as in (\ref{-+}) we recast the above
equation into
\begin{equation}
\label{mean-2-simple} 
\frac{d}{dt}\,\langle
x^2\rangle=\frac{1}{2}\sum_{n=-\infty}^\infty(P_n+Q_n)=\frac{1}{2}
\end{equation}
The last identity is implied by normalization and its validity also
follows from Eqs.~\eqref{Pn}--\eqref{Qn}. Integrating
\eqref{mean-2-simple} yields  $\langle x^2\rangle=\frac{1}{2}\,t$
which in conjunction with (\ref{D-extract}) leads to the previously
announced result, Eq.~\eqref{D2}.

\subsection{General Case}
\label{two-leg} 

In the general case ($s\geq 2$) we denote
\begin{equation}
\label{Pnl}
P_n^\ell={\rm Prob}[\bullet\underbrace{\circ\cdots\circ}_{\ell-1}\bullet]
\end{equation}
the probability to occupy sites $n$ and $n+\ell$. These probabilities
obey
\begin{subequations}
\begin{align}
&\frac{d P_n^{1}}{dt}=P_{n-1}^{2}+P_{n}^{2}-2P_n^{1}
\label{Pn1}\\
&\frac{d P_n^{\ell}}{dt}
=P_{n-1}^{\ell+1}+P_{n+1}^{\ell-1}+P_{n}^{\ell+1}+P_{n}^{\ell-1}-4P_n^{\ell}
\label{Pns}\\
&\frac{d P_n^{s}}{dt}=P_{n+1}^{s-1}+P_{n}^{s-1}-2P_n^{s}
\label{PnL}
\end{align}
\end{subequations}
where equations \eqref{Pns} apply for $2\leq \ell\leq s-1$. The
mean-square displacement is given by
\begin{equation}
\label{mean-s-def} 
\langle x^2\rangle=\sum_{n=-\infty}^\infty\sum_{\ell=1}^s
\left(n+\frac{\ell}{2}\right)^2 P_n^{\ell}
\end{equation}
Using Eqs.~\eqref{Pn1}--\eqref{PnL} and applying the same tricks as
above to simplify the sums, we obtain 
\begin{equation}
\label{mean-s} 
\frac{d}{dt}\,\langle x^2\rangle =\sum_{\ell=1}^s w_\ell
-\frac{1}{2}\left(w_1+w_s\right)
\end{equation}
where $w_\ell=\sum_{n} P_n^{\ell}$ is the weight of configurations of
the type  $\mathcal{C}_\ell$.  The sum on the right-hand side of
Eq.~\eqref{mean-s} is equal to one due to normalization.  To determine
$w_1$ and $w_s$ one does not need to solve an infinite set of the
master equations \eqref{Pn1}--\eqref{PnL}. Instead, we take
Eqs.~(\ref{Pn1})--(\ref{PnL}) and sum each of them over all $n$ to
yield a closed system of equations for the weights
\begin{subequations}
\begin{align}
&\frac{d w_1}{dt}=2(w_2-w_1)
\label{w1}\\
&\frac{d w_\ell}{dt}=2(w_{\ell-1}+w_{\ell+1}-2w_\ell)
\label{ws}\\
&\frac{d w_s}{dt}=2(w_{s-1}-w_s)
\label{wL}
\end{align}
\end{subequations}
If initially $w_1=\ldots=w_s=1/s$, then Eqs.~(\ref{w1})--(\ref{wL})
show that this remains valid forever.  Even if we start with an
arbitrary initial condition, all the weights $w_s$   relax exponentially
 fast  toward the `equilibrium' value $1/s$. Thus the
right-hand side of (\ref{mean-s}) becomes $1-1/s$ yielding $\langle
x^2\rangle=(1-1/s)\,t$ which in conjunction with (\ref{D-extract})
leads to Eq.~\eqref{Ds}.

\subsection{Heterogeneous Spiders}
\label{lame}

Various spiders can be assembled experimentally \cite{exp}, including those with distinguishable legs. Here we analyze the coarse-grained properties of these `lame'
spiders.

The bipedal lame spider is characterized by the maximal separation $s$
between the legs and  by the hopping rates $\alpha$ and $\beta$ of the
legs, e.g.,  the $\alpha$-leg hops to the right and left with the
same rate $\alpha$ (whenever hopping is possible) in the symmetric
case. For the bipedal spider with $s=2$, the diffusion coefficient is
given by 
\begin{equation}
\label{DAB} 
D_2=\frac{1}{2}\,\frac{\alpha\beta}{\alpha+\beta}
\end{equation}
When $\alpha=\beta$ we recover the already known result telling us
that the diffusion coefficient is 4 times smaller than the hopping
rate. For a  very lame spider  ($\alpha\ll \beta$), Eq.~\eqref{DAB}
gives $D_2=\alpha/2$, so the diffusion coefficient is half 
 the hopping rate of the very slow leg.

To derive \eqref{DAB}, we first note that the probabilities
\eqref{PQ-def} satisfy
\begin{subequations}
\begin{align}
&\frac{d P_n}{dt} = \beta Q_{n} + \alpha Q_{n-1}-(\alpha+\beta)P_n
\label{Pn-lame}\\
&\frac{d Q_n}{dt}=\alpha P_{n+1} + \beta P_{n}-(\alpha+\beta)Q_n
\label{Qn-lame}
\end{align}
\end{subequations}
Here we have assumed that the left leg hops with rate  $\alpha$ and
the right leg hops with rate  $\beta$. (Recall that in one dimension,
the order of the legs never changes.)

Using Eqs.~\eqref{Pn-lame}--\eqref{Qn-lame} we find that the
mean-square displacement \eqref{mean-2-simple} evolves according to
\begin{equation}
\label{mean2-lame} 
\frac{d}{dt}\,\langle x^2\rangle=\frac{\alpha+\beta}{4} -
(\alpha-\beta)(u-v)
\end{equation}
where
\begin{equation*}
u=\sum_{n=-\infty}^\infty \left(n+\frac{1}{2}\right)P_n\,,\quad
v=\sum_{n=-\infty}^\infty (n+1)Q_n
\end{equation*}
{}From Eqs.~\eqref{Pn-lame}--\eqref{Qn-lame} we deduce that the quantity $u-v$ obeys
\begin{equation}
\label{uv-eq} 
\frac{d}{dt}\,(u-v)=\frac{\alpha-\beta}{2}-2(\alpha+\beta)(u-v)
\end{equation}
Equation (\ref{uv-eq}) shows that $u-v$ quickly approaches to
$(\alpha-\beta)/[4(\alpha+\beta)]$. Plugging this into
(\ref{mean2-lame}) yields
\begin{equation*} 
\frac{\langle x^2\rangle}{t}\to
\frac{1}{4}\left[\alpha+\beta-\frac{(\alpha-\beta)^2}{\alpha+\beta}\right]
=\frac{\alpha\beta}{\alpha+\beta}
\end{equation*}
which leads to the announced result, Eq.~\eqref{DAB}.

For the biased bipedal lame spider, the drift velocity is given by a
neat formula
\begin{equation}
\label{V-lame}
V_2 = \frac{\alpha\beta}{\alpha+\beta}
\end{equation}
which resembles to \eqref{DAB}. To establish \eqref{V-lame} one can use
the analog of  Eqs.~\eqref{Pn-lame}--\eqref{Pn-lame}, namely
\begin{subequations}
\begin{align}
&\frac{d P_n}{dt} = \alpha Q_{n-1} - \beta P_n
\label{Pn-lame-bias}\\
&\frac{d Q_n}{dt} = \beta P_{n} - \alpha Q_n
\label{Qn-lame-bias}
\end{align}
\end{subequations}
Equations \eqref{Pn-lame-bias}--\eqref{Qn-lame-bias} give the weights
\begin{equation*}
w_1\equiv \sum_{n=-\infty}^\infty P_n =
\frac{\alpha}{\alpha+\beta}\,,\quad w_2\equiv \sum_{n=-\infty}^\infty
Q_n = \frac{\beta}{\alpha+\beta}
\end{equation*}
and relation $V=(\beta w_1 + \alpha w_2)/2$ leads to \eqref{V-lame}.

Further analysis of Eqs.~\eqref{Pn-lame-bias}--\eqref{Qn-lame-bias}
allows one to determine the diffusion coefficient
\begin{equation}
\label{D-lame} 
D_2 = \frac{1}{2}\,\alpha\beta\,\frac{\alpha^2 +
\beta^2}{(\alpha+\beta)^3}
\end{equation}
We do not give a derivation of this formula since it can be extracted
from earlier results by Fisher and Kolomeisky  \cite{FK} who in turn
used previous findings by  Derrida~\cite{D}.

It is more difficult to compute the diffusion coefficient for the
bipedal lame spider with maximal span $s>2$. The results of
Ref.~\cite{FK} do not cover the general case, although a proper
extension of methods \cite{D,FK} may solve the problem.
 For the symmetric  bipedal lame spider with maximal span $s\ge 2$,  we used an approach outlined in Appendix~\ref{AppA}   and obtained
\begin{equation}
\label{DAB-s} 
D_s=\frac{\alpha\beta}{\alpha+\beta}\,\left(1-\frac{1}{s}\right). 
\end{equation}
For $s=2$, we recover equation~\eqref{DAB}.

For the biased bipedal lame spider with maximal span $s \ge 2$ it is again
simple to determine the drift velocity. Using an analog of
\eqref{Pn-lame-bias}--\eqref{Qn-lame-bias} one gets the weights and
then the drift velocity is found from the relation
\begin{equation*}
2V_s = \beta w_1+(\alpha+\beta)\sum_{\ell=2}^{s-1}w_\ell + \alpha w_s
\end{equation*}
The outcome of this computation is
\begin{equation}
\label{V-lame-gen}
V_s = \alpha\beta\,\,\frac{\alpha^{s-1} - \beta^{s-1}}{\alpha^s -
\beta^s}
\end{equation}
Specializing to $\alpha=\beta=1$ (the l'Hospital rule allows to
resolve an apparent singularity) one arrives at the expression  \eqref{VDL} 
 for the velocity.

Finally, the diffusion coefficient for the biased bipedal spider with
arbitrary $s$ is
\begin{eqnarray}
\label{Ds-lame}
D_s &=& \frac{1}{2}\,\alpha\beta\,\,\frac{\alpha^{s-1} -
\beta^{s-1}}{\alpha^s - \beta^s}\nonumber\\
&+&\frac{1+s}{\alpha^s - \beta^s}
    \,\,\frac{\alpha^{s-1} - \beta^{s-1}}{\alpha^s - \beta^s}\,\,
\frac{\alpha^{s+1}\beta^{s+1}}{\alpha^s - \beta^s}\nonumber\\
&+& \frac{1-s}{\alpha^s - \beta^s} \,\, 
\frac{\alpha^{s+1} - \beta^{s+1}}{\alpha^s - \beta^s} \,\,
 \frac{\alpha^s \beta^s}{\alpha^s - \beta^s} \, . 
\end{eqnarray}
This equation is derived in Appendix~\ref{AppA}.

Equation \eqref{Ds-lame} reduces to \eqref{D-lame} when $s=2$; for $s=3$, the diffusion coefficient can be re-written as
\begin{equation*}
D_3 = \frac{1}{2}\,\alpha\beta\,
\frac{(\alpha+\beta)(\alpha^2 - \alpha\beta +\beta^2)(\alpha^2 +3\alpha\beta +\beta^2)}{(\alpha^2 + \alpha\beta +\beta^2)^3}
\end{equation*}
Also when $\alpha=\beta=1$, equation \eqref{Ds-lame} reduces to the
expression \eqref{VDL} for the  diffusion coefficient.

\section{Multi-pedal spiders}
\label{multi}

For the multi-pedal spider, $L\geq 3$, we must specify the constraint
governing the separations between the legs.

\begin{figure}
\centering
\includegraphics[scale=0.13]{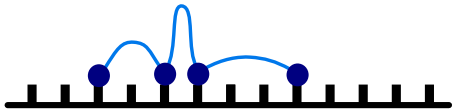}
\caption{Illustration of a centipede, i.e. a spider with local constraint. All legs step to empty nearest neighbor sites at the same rate with adjacent legs staying within distance $s$ from each other.}
\label{local}
\end{figure}

\subsection{Centipedes}

Here we consider centipedes or local spiders where the distance between the $j^{\rm th}$ and $(j+1)^{\rm st}$ legs is at most $s$. See Fig.~\ref{local} for an illustration of such a centipede. In this case the total number of configurations is
$\mathcal{C}=s^{L-1}$ since each of the $(L-1)$ spacings between
adjacent legs can have $s$ possible values.

\subsubsection{Main results}
\label{main}

Consider first spiders with $s=2$. The configurations for the bipedal spider are shown in \eqref{2}; the four possible configurations for
the tripod are
\begin{equation}
\label{3} 
\bullet\bullet\bullet \qquad \bullet\circ\bullet\bullet\qquad
\bullet\bullet\circ\bullet\qquad \bullet\circ\bullet\circ\, \bullet
\end{equation}
and generally there are $2^{L-1}$ possible configurations.

Let $D(L)$ be the diffusion coefficient of an $L$-leg spider. In the
case of symmetric hopping (all rates are one)
\begin{equation}
\label{DN} 
D(L)=\frac{1}{4(L-1)}
\end{equation}
when $s=2$. For $L=2$, this of course agrees with our previous result:
$D(2)=D_2=1/4$.

For the biased multi-pedal spider, the velocity is
\begin{equation}
\label{VL}
V(L)=\frac{1}{2}\,\frac{L+1}{2L-1}
\end{equation}
The biased infinite-leg spider has a finite limiting speed! More
precisely, $V(\infty)=1/4$, i.e.,  the infinite-leg spider drifts 4
times slower than the single-leg spider. The diffusion coefficient of
the biased spider is
\begin{equation}
\label{DL}
D(L)=\frac{3}{4}\,\frac{(4L-3)!\, [(L-1)! (L+1)!]^2}{[(2L-1)!]^3\,
(2L+1)!}
\end{equation}
Note that the diffusion coefficient of the infinite-leg spider
vanishes. Asymptotically,
\begin{equation}
\label{DL-largeL}
D(L) \sim \frac{3\sqrt{2\pi}}{128}\,L^{-1/2}\quad {\rm as}\quad
L\to\infty
\end{equation}

The above results \eqref{DN}--\eqref{DL} are valid when $s=2$. We
have not succeeded in computing $V(L)$ and $D(L)$ for arbitrary $L$
when the maximal separation exceeds two, $s>2$. 

The velocity and the diffusion coefficient can be computed for centipedes with $s>2$ when the number of legs is sufficiently small. The simplest quantity is the velocity of biased spiders. When $s=3$, we computed the velocity $V(L)$  of centipedes with up to seven legs: 
\begin{equation*}
\begin{split}
 V(2) &=  2/3\\
 V(3) &= 26/45 \approx 0.5778\\
 V(4) &= 2306/4301\approx 0.5362\\
 V(5) &= \frac{2257932864491452}{4410656468591479} \approx 0.5119\\
 V(6) 
&\approx 0.4960476429\\
V(7) &\approx 0.4848259795\\
\end{split}
\end{equation*}
(we have not displayed exact expressions for $V(6)$ and $V(7)$ which are the ratios of huge integers.) Note that for biased spiders with $s=3$ one can guess the general expression \eqref{VL} from exact results for $V(L)$ for small $L$; in contrast, 
no simple expression seems to exist for the velocity of biased spiders with $s>2$. 

For symmetrically hopping spiders, we computed the diffusion coefficient when the number of legs is small. Here are the results for centipedes with $s=3$ (the method used in calculations is described in Appendix \ref{AppA})
\begin{equation*}
\begin{split}
 D(2) &=  1/3\\
 D(3) &= 22/117 \approx 0.1880\\
 D(4) &= 530/4059 \approx 0.1306\\
 D(5) &= \frac{145730406362990}{1457669284934841} \approx 0.0999749\\
 D(6)
 &= \frac{13157424169190800305558220463956878370565}{162454344889141072641777603974162004103911}\\
 &\approx 0.080991519
\end{split}
\end{equation*}
In contrast to the neat formula \eqref{DN} characterizing the
$s=2$ case, the above numbers  look intimidating. Factorizing the
nominator and denominator of $D(6)$ reveals the presence of
extraordinary huge factors and thereby excludes that it can be
described by a formula like \eqref{DL}, let alone \eqref{DN}.  Note
that at least the $L^{-1}$ asymptotic behavior predicted by
Eq.~\eqref{DN} remains valid for all $s$; for $s=3$, in particular, we
have $D(L)\sim AL^{-1}$ with  $A\approx 0.423$ when $L\gg 1$.

\subsubsection{Mapping to the exclusion process for $s = 2$}

The derivations of above results are complicated since the number of
configurations grows exponentially with $L$. Further, the transition
rates are configuration dependent, e.g., for the four-leg spider
configurations
\begin{equation*}
\bullet\bullet\bullet\bullet\qquad
\bullet\circ\bullet\bullet\bullet\qquad
\bullet\bullet\circ\bullet\, \bullet
\end{equation*}
evolve with rates 2, 3, 4 for symmetric hopping. (In contrast, for
bipedal spiders the number of configurations grows linearly with $s$
and the transition rates are simple.)  All this makes the computation
of the diffusion coefficient $D(L)$ for arbitrary  $L$ very
challenging.  The pedestrian calculation is feasible for small $L$,
but even for $L=3$, the framework based on rate equations like
\eqref{Pn1}--\eqref{PnL} is very cumbersome.

Fortunately, spiders with local constraint and $s=2$ are related to simple
exclusion processes (SEPs). This allows us to extract some predictions about spiders from previously known results about SEPs, and to employ the methods developed in the context of SEPs to situations natural in applications to spiders. 

We now demonstrate the remarkable connection between centipedes with $s=2$ and SEPs. As an example we show that the  biased spider is isomorphic to the totally asymmetric simple exclusion process (TASEP)
with open boundary conditions. To understand the isomorphism, consider
for concreteness the tripod. We can map configurations \eqref{3} onto
configurations
\begin{equation}
\label{3-map}
00\qquad 10\qquad 01\qquad 11
\end{equation}
of the exclusion process on two sites with open boundary conditions.
Here $0$ on $j^{\rm th}$ site implies that there is no empty site
between  $j^{\rm th}$ and $(j+1)^{\rm st}$ legs, while $1$ implies that
there is an empty site. A  hop to the right of an internal leg in
\eqref{3} corresponds to a  hop to the left of a particle in
\eqref{3-map}. Further, the hop of an extreme right leg corresponds to
the addition of a particle to the extreme right position, and the hop
of the extreme left leg corresponds to the removal of a particle from
the extreme left position. The same mapping applies to any $L$. Thus
in this TASEP each site $i=1,\ldots,L-1$ can be occupied by a
particle, and each particle hops to the left with rate one if this
site is empty; further, a particle is added to site $i=L-1$ with rate
one if this site is empty, and a particle is removed from site $1$
with rate one if this site is occupied. Thus we have shown that the  $s=2$   biased spider  that moves to the right is equivalent to the  TASEP with open boundaries in which particles hop from right to left. A similar mapping holds  between the symmetric  spider  and the symmetric exclusion process. 

Derrida, Domany, and Mukamel \cite{DDM} have shown that Eq.~\eqref{VL} gives the flux in the TASEP; the isomorphism between the flux and velocity proves that the velocity of the biased spiders is given by \eqref{VL}. This result was
re-derived by other techniques, e.g.,  by a pure combinatorial approach
\cite{gilles}.  The (much more complicated) derivation of the diffusion coefficient
in Ref.~\cite{DEM} gives  \eqref{DL}.

\subsubsection{Derivation of \eqref{DN}}

For the  symmetric  spider, it should be possible to compute the diffusion coefficient  \eqref{DN} by using the technique of Ref.~\cite{DEM}. This technique (based on an extension of a matrix technique) is very advanced. The final result \eqref{DN} looks much simpler than its biased counterpart \eqref{DL}. Hence we have sought another derivation, and we have found an intriguingly simple proof of \eqref{DN} based on the fluctuation-dissipation formula (see Appendix \ref{AppB}).

First we recall that the symmetric  spider with $L$ legs hopping in both directions with   rates equal to 1 is equivalent to a symmetric  exclusion process on $L-1$ sites with open boundary conditions. For this SEP, {\it all}  rates (i.e. hopping rates in the bulk, entrance and exit rates at  the boundaries)  are equal to 1. This  Markov process  satisfies detailed balance and is at equilibrium; in particular, the mean current, i.e.,  the velocity of the spider, vanishes identically.  The variance of the current corresponds to the diffusion constant of the spider. This variance can be calculated
as follows.

Consider now a symmetric exclusion process of length $L-1$ with open boundaries and {\em arbitrary} addition and removal rates at the boundaries. The system is driven out of equilibrium by particles entering and leaving  at the boundaries.  In the bulk, each particle hops with rates 1 to the right and to the left (if the corresponding sites are empty);  a particle enters at site 1 with rate 
$\alpha$ and leaves this site with rate $\gamma$; similarly a particle enters another boundary site $L-1$ with rate $\delta$ and leaves this site  with rate $\beta$. Generically,  these unequal rates  lead to a current. The mean value of this current is given by (see e.g. Ref.~\cite{DPramana})
\begin{equation}
 J = \frac{ \frac{\beta}{\beta + \delta} -  \frac{\gamma}{\alpha+\gamma} }
  { L  +  \frac{1}{\alpha +\gamma} +  \frac{1}{\beta + \delta} -2 } \, .
 \label{Jsym} 
\end{equation}
 The equilibrium conditions correspond to $\alpha = \gamma = \beta =\delta
 = 1$ and $J =0$.  We now  choose the rates  on site 1 as follows
  $\alpha =  \exp(\frac{\epsilon}{2})$ and   $\gamma =
  \exp(-\frac{\epsilon}{2})$ and   we keep 
   $\beta =\delta = 1$ at site $L-1$. Then, 
 the current is given by 
\begin{equation}
 J =  \frac{\tanh\frac{\epsilon}{2}}
 {2L -  3  + \frac{1}{ \cosh\frac{\epsilon}{2}}} \, . 
\end{equation}

The Markov matrix of  this process satisfies the generalized detailed balance condition given by  equation \eqref{gendetbal} of  the Appendix  \ref{AppB}, with  
$y =\pm 1$ if a particle enters at site 1, or exits from site 1 ($y = 0$ otherwise).
We can then use the fluctuation-dissipation formula (see \eqref{eq:FDT} in the Appendix  \ref{AppB}) wich  tells us that the fluctuation of the  current at the first site 
is given by 
\begin{equation}
 D = 
  \frac{ \partial J}{ \partial\epsilon} \Big|_{\epsilon=0}= \frac{1}{4(L-1)}
 \, , 
\end{equation}
in accordance with equation  \eqref{DN}.

\subsubsection{Mean-field approximation for $s \ge 3$}

Simple exclusion processes have been thoroughly investigated (see
books and reviews \cite{spohn,derrida,liggett,gunter}). Hence one can
extract the results about spiders from already known results about
SEP. Unfortunately, for spiders with local constraint the mapping onto
SEP applies only when $s=2$. The spider with $L$ legs and arbitrary
$s$ can be mapped onto an exclusion-like process with 
$L-1$ sites and  with open boundaries. In this process the maximal
occupancy is limited, namely the number of particles in each site
cannot exceed $s-1$. The dynamics is simple: one chooses sites with
rate one and moves a particle to the site on its left; nothing happens if
the chosen site was empty or the site on the left was fully
occupied. One also adds particles to site $i=L-1$ and removes from
site $i=1$, both these processes occur with rate one; the addition is
possible as long as site $i=L-1$ is not fully occupied (contains no
more than $s-1$ particles). Unfortunately, this neat process has not
been solved exactly but it can be studied  by a mean-field analysis.

To simplify the analysis, we consider centipedes with infinitely many legs. We assume that the distance between adjacent legs cannot exceed $s$. We further assume that the spider's motion is biased, and limit ourselves to a (mean-field) computation of its velocity $V^{(s)}$.

First we map the spider onto the generalized asymmetric exclusion process with at most $s-1$ particles per site. We then write $x_j$ for the density of sites with $j$ particles; this is just the density of gaps of length $j+1$ between adjacent legs of the spider. The possible values are $j=0,\ldots, s-1$. Writing the evolution equation for $\dot x_j$ and setting $\dot x_j=0$ we obtain
\begin{equation}
\label{xj}
(x_{j-1}-x_j)(1-x_0)-(x_j-x_{j+1})(1-x_{s-1})=0
\end{equation}
when $1\leq j\leq s-2$. Similarly from $\dot x_0=0$ and $\dot x_{s-1}=0$  we get
\begin{subequations}
\begin{align}
&x_1(1-x_{s-1})-x_0(1-x_0)=0
\label{x0}\\
&x_{s-2}(1-x_0)-x_{s-1}(1-x_{s-1})=0
\label{xs}
\end{align}
\end{subequations}
The obvious normalization requirement is
\begin{equation}
\label{conserv}
\sum_{j=0}^{s-1}x_j=1
\end{equation}

As a warm up, consider the first non-trivial case $s=3$. Due to normalization, it is sufficient to use 
\eqref{x0}--\eqref{xs}. Writing $x_0\equiv x$ and $x_2\equiv z$, we have 
$x_1=1-x-z$ from \eqref{conserv}, and \eqref{x0}--\eqref{xs} become
\begin{subequations}
\begin{align}
&(1-x-z)(1-z)=x(1-x)
\label{x}\\
&(1-x-z)(1-x)=z(1-z)
\label{z}
\end{align}
\end{subequations}
These equations are actually identical; solving any of them we arrive at
\begin{equation}
\label{x-sol}
x=1-\frac{z+\sqrt{z(4-3z)}}{2}
\end{equation}

To compute velocity we return to the original formulation. A leg of the spider moves with rate 1 if the site ahead is empty and if the leg behind is one or two steps behind. The former event happens with probability $1-x$, while the latter occurs with probability $1-z$. Thus the velocity is 
\begin{equation}
\label{Vxz}
V^{(3)}=(1-x)(1-z)
\end{equation}
Using \eqref{x-sol} we get 
\begin{equation}
\label{Vz}
V=\frac{1}{2}\,(1-z)\left[z+\sqrt{z(4-3z)}\right]
\end{equation}
We should select maximal velocity. The maximum of $V(z)$ given by \eqref{Vz} is reached at $z=1/3$, and it reads
\begin{equation}
\label{V-s3}
V^{(3)}=\frac{4}{9}
\end{equation}
At the state corresponding to the actual (maximal) velocity all
densities are equal: $x_0=x_1=x_2=1/3$.   

The situation for $s>3$ is also simple. Analyzing recurrence \eqref{xj} one finds that for all $0\leq j\leq s-1$ the solution is a shifted geometric progression
\begin{equation}
\label{xAB}
x_j=A+B\lambda^j, \qquad \lambda=\frac{1-x_0}{1-x_{s-1}}
\end{equation}
Plugging \eqref{xAB} into Eqs.~\eqref{x0}--\eqref{xs} one achieves the consistency if either $A=0$ or $\lambda=1$. In the latter case the densities are the same, and hence they are all equal to $s^{-1}$ due to normalization requirement  \eqref{conserv}. The straightforward generalization of \eqref{Vxz} is 
\begin{equation}
\label{Vs-eq}
V^{(s)}=(1-x_0)(1-x_{s-1})
\end{equation}
and therefore
\begin{equation}
\label{V-s}
V^{(s)}=\left(1-\frac{1}{s}\right)^2
\end{equation}

In the complimentary case of $A=0$ the analysis is a bit more lengthy. However, the final result is the same. Here is the proof. Since $x_j=B\lambda^j$, equation \eqref{xAB} gives $\lambda=(1-B)/(1-B\lambda^{s-1})$, which can be re-written as
\begin{equation}
\label{B-eq}
B=\frac{1-\lambda}{1-\lambda^s}
\end{equation}
Further, \eqref{Vs-eq} becomes
\begin{equation}
\label{Vs-B}
V^{(s)}=(1-B)(1-B\lambda^{s-1})
\end{equation}
Using \eqref{B-eq} we recast \eqref{Vs-B} into 
\begin{equation}
\label{Vs-lambda}
V^{(s)}(\lambda) = \lambda\left(\frac{1-\lambda^{s-1}}{1-\lambda^s}\right)^2
\end{equation}
The maximum of $V^{(s)}(\lambda)$ is achieved at $\lambda=1$. Thus the velocity is indeed given by Eq.~\eqref{V-s}.

The above elementary analysis is mean-field as we have assumed the validity of the factorization. The answer is trivially exact for $s=1$, and it is known to be exact for 
$s=2$.  For $s=3$, we calculated  velocities exactly for small centipedes, see  
Sect.~\ref{main}. The limiting $L\to\infty$ value obtained from simulations $V^{(3)}\approx 0.4189$ is close to the predicted mean-field value $V^{(3)}=4/9\approx 0.4444$. Overall, the assumed factorization is not exact when $s\geq 3$. Note that a model which differs from our model only in the hopping rules has been solved exactly \cite{gunter2}, but there the stationary state is a product measure.

\subsubsection{Lame spiders}

Finally we investigate lame centipede spiders whose extreme left leg hops to the right with rate $\alpha$ and the extreme right leg hops to the right with rate $\beta$. The above mean-field analysis shows that the velocity of the extreme left leg is $\alpha(1-x_0)$ and the velocity of the extreme right leg is $\beta(1-x_{s-1})$. As long as these velocities exceed the bulk velocity \eqref{V-s}, the actual gap density $x_0$ at the left end and $x_{s-1}$ at the right end will be higher than their bulk values, so the spider will move with velocity \eqref{V-s}. This occurs as long as $\alpha(1-s^{-1})$ and 
$\beta(1-s^{-1})$ exceed $(1-s^{-1})^2$, i.e. $\alpha,\beta\geq 1-s^{-1}$. When at least one of the rates is smaller than the threshold value, different behaviors emerge. Overall, the speed of the infinite-leg spider exhibits an amusing dependence on the rates 
$\alpha$ and $\beta$:
\begin{equation}
\label{V-inf-s}
V^{(s)} = \left\{   
\begin{array}{cl} \displaystyle
(1-s^{-1})^2 & \mbox{~~for~~} \alpha,\beta\geq 1-s^{-1}\\ 
\displaystyle
W_s(\alpha) &  \mbox{~~for~~} \alpha\leq\beta,~ \alpha<1-s^{-1}\\ 
\displaystyle
W_s(\beta)  &  \mbox{~~for~~} \beta\leq\alpha,~ \beta<1-s^{-1}
\end{array}
\right.
\end{equation}
Thus if at least one of the two extreme legs has the intrinsic speed less than $1-s^{-1}$, the speed of the entire spider is solely determined by the slowest leg. 

To determine $W_s(\beta)$ we note that velocity on the right boundary is
\begin{equation}
\label{V-beta}
V=\beta(1-x_{s-1})=\beta(1-B\lambda^{s-1})=\beta\,\frac{1-\lambda^{s-1}}{1-\lambda^s}
\end{equation}
where in the last step we have used \eqref{B-eq}. Equating the velocity given by 
Eq.~\eqref{V-beta} with the velocity in the bulk given by
Eq.~\eqref{Vs-lambda}  we find
\begin{equation}
\label{beta-eq}
\beta=\lambda\,\frac{1-\lambda^{s-1}}{1-\lambda^s}
\end{equation}
Thus the velocity is given by \eqref{V-beta} or \eqref{Vs-lambda}, where parameters are connected via \eqref{beta-eq}. 

Explicit results can be obtained for $s$ up to $s=5$. For $s=2$ we recover the celebrated result 
\begin{equation}
\label{W2}
W_2(\beta)=\beta(1-\beta)
\end{equation}
For $s=3$ the final expression is still compact
\begin{equation}
\label{W3}
W_3(\beta)=\beta(1-\beta)\,\frac{1+\sqrt{1+4b}}{2}
\end{equation}
with $b=\beta/(1-\beta)$. For $s=4$ the result is quite cumbersome
\begin{equation}
\label{W4}
W_4(\beta)=\frac{\beta^2}{\lambda}\,,\quad 
\lambda=\frac{1}{6}\,\Delta-\frac{4}{3}\,\Delta^{-1}-\frac{1}{3}
\end{equation}
where we have used the shorthand notation
\begin{eqnarray*}
\Delta=\left\{28+108 b+12
\sqrt{9+42 b+81 b^2}\right\}^{1/3}
\end{eqnarray*}

\subsection{Global Constraint}

\begin{figure}
\centering
\includegraphics[scale=0.13]{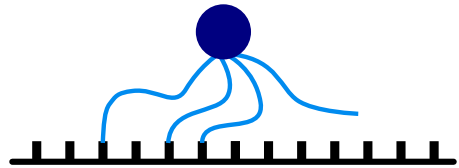}
\caption{Illustration of a spider with global constraint. The legs can step independently to nearest neighbor empty sites within a distance $S$ from each other.}
\label{global}
\end{figure}

For spiders with $L$ legs and maximal span $S$ between any two legs (see Fig.~\ref{global}), the global constraint rule limits the maximal distance between the extreme legs and the exclusion condition implies that $S\geq L-1$. A spider with maximal distance $S=L-1$ is immobile, so we shall tacitly assume that $S\geq L$. It is also useful to keep in mind that for a spider satisfying the local constraint rule the maximal span is $(L-1)s$ if the maximal distance between the adjacent legs is $s$; for the bipedal spider $S\equiv s$.  A spider  with global constraint is equivalent  to the exclusion  process on  a {\em  ring}, where each leg is interpreted as a particle and the total number of sites is equal to $S+1$. For such a process with periodic boundary conditions,  a  key property of the stationary state, which holds both in symmetric and biased cases,  is that all configurations have equal weight \cite{derrida}.

\subsubsection{Configurations}

To count the total number of configurations, we set, as usual, the origin at the
position of the extreme left leg,  see e.g.,  \eqref{3};
this allows us to avoid multiple counting of configurations which
differ merely by translation. We then note that the other $L-1$ legs
can occupy sites $1,\ldots,S$. Thus the total number of configurations
is
\begin{equation}
\label{number-conf}
\mathcal{C}(L,S)=\binom{S}{L-1}
\end{equation}
 In the stationary state, the   weight of a configuration is thus
 given by   $w=1/\mathcal{C}$.

Let us now calculate  the total number $\mathcal{N}(L,S)$ of
$\bullet\circ$ pairs in  {\em all} configurations. Each configuration
begins with a string
\begin{equation}
\label{string}
\bullet\underbrace{\circ\cdots\circ}_{a}\bullet
\end{equation}
where $a=0,1,\ldots,S-L+1$.  Disregarding the part up to the second
leg  maps configurations of the type \eqref{string} with fixed $a$ to
configurations of the spider with $L-1$ legs and maximal span
$S-a$. The total number of $\bullet\circ$ pairs in these latter
configurations is $\mathcal{N}(L-1,S-a)$. Configurations of the type
\eqref{string} have of course an additional $\bullet\circ$ pair at the
beginning  (when $a>0$). Therefore
\begin{equation*}
\mathcal{N}(L,S)=\sum_{a=0}^{S-L+1}\mathcal{N}(L-1,S-a)+
\sum_{a=1}^{S-L+1}\mathcal{C}(L-1,S-a)
\end{equation*}
The latter sum is simplified by using \eqref{number-conf} and the  identity
\begin{equation*}
\sum_{p=q}^r \binom{p}{q} = \binom{r+1}{q+1}
\end{equation*}
Thus we arrive at the recurrence
\begin{equation}
\label{NLS-rec}
\mathcal{N}(L,S) = \sum_{a=0}^{S-L+1}\mathcal{N}(L-1,S-a)+
\binom{S-1}{L-2}
\end{equation}
The solution (found by a generating function technique or verified by
mathematical induction) reads
\begin{equation}
\label{NLS}
\mathcal{N}(L,S) = L\binom{S-1}{L-1}
\end{equation}

Since all configurations have equal weight in the stationary state, the velocity of the biased spider can be expressed by the total number  $\mathcal{N}$ of $\bullet\circ$ pairs as
\begin{equation}
 V = L^{-1}\, \frac{\mathcal{N}}{\mathcal{C}}\, ,
 \label{V-Tibor}
\end{equation}
which then leads to
\begin{equation}
\label{V-global}
V=1-\frac{L-1}{S}
\end{equation}
(Note that the velocity is zero in the unbiased case.) It is more involved to calculate the diffusion coefficient, which we obtain below separately for the unbiased and for the biased case.

\subsubsection{Symmetric hopping}

 The diffusion coefficient  of the symmetric spider is given by
\begin{equation}
\label{D-Tibor}
D=L^{-2}\,\frac{\mathcal{N}}{\mathcal{C}} \, .
\end{equation}
 This equation is obtained by applying the fluctuation-dissipation
 relation, which is valid because the  dynamics of the 
  symmetric spider statisfies
 detailed balance (in other words, the  symmetric spider 
 is a system  in thermodynamic equilibrium). The Einstein relation
 then implies that  $D \propto V$, where  the velocity
 $V$  of the biased spider is given by 
 the  expression~\eqref{V-Tibor}.  The extra factor $L^{-1}$
 between \eqref{D-Tibor} and~\eqref{V-Tibor} comes from the fact that
 $D$ is the diffusion coefficient of the left leg (or of the spider's
 center of mass) whereas $V$ is the mobility of the biased spider 
 where all the legs are asymmetric. 
Using Eqs.~\eqref{number-conf} and
\eqref{NLS} we recast  \eqref{D-Tibor} to
\begin{equation}
\label{D-global}
D=\frac{1}{L}\left[1-\frac{L-1}{S}\right] 
\end{equation}
For the bipedal spider there is no difference between local and global
constraints. Using $L=2$ and $S=s$ we find that Eq.~\eqref{D-global}
indeed turns into Eq.~\eqref{Ds}. The  calculation  of $D$ presented
 here is self-contained;  we notice that  the expression~(\ref{D-global})
  can also be found   as a special 
 case of  a general formula  derived in \cite{DMal} 
  for the diffusion constant of a partially
 asymmetric exclusion process. 

\subsubsection{Biased hopping}

While the velocity \eqref{V-global} is easily computable, the diffusion
coefficient was calculated in \cite{DE} 
 using a matrix Ansatz
  (see \cite{derrida} and  \cite{DMal}  for a more general formula).
  The result is
\begin{equation}
\label{DL-ring}
D(L,S)=\frac{1}{2(2S-2L+1)}\,\, \binom{2S-1}{2L-1}\,\binom{S}{L-1}^{-2}
\end{equation}

For a given number of legs, the diffusion coefficient of the most
clumsy spider is
\begin{equation}
\label{DL-clumsy}
D(L,L)=\frac{1}{2 L^2}
\end{equation}
while the diffusion coefficient of the most agile spider is
\begin{equation}
\label{DL-agile}
D(L,\infty)=\frac{2^{2L-2}}{L}\,\, \binom{2L}{L}^{-1}
\end{equation}
When $L\gg 1$, the diffusion coefficient \eqref{DL-agile} scales as
\begin{equation}
\label{DL-large}
D(L,\infty)\sim \frac{1}{4}\,\sqrt{\frac{\pi}{L}}
\end{equation}
More generally, the diffusion coefficient $D(L,S)$ also  decreases as
$(\pi/16 L)^{1/2}$ when $1\ll L\ll S^{1/2}$.

\subsection{Heterogeneous Spiders}

Each leg of a heterogeneous (lame) spider may have its own hopping
rate. The bipedal lame spider was studied in Sec.~\ref{lame}. One can
find explicit expressions for the velocity and the diffusion
coefficient of the lame tripod and perhaps for the lame spider with
four legs; the general solution for an arbitrary $L$ is unknown.

Lame spiders are tractable if only one or two legs have different
hopping rates. Below we consider lame spiders whose extreme legs are
affected. For concreteness, we focus on spiders with local constraint
and $s=2$. The analogy with the TASEP with open boundary conditions
still applies, the only modification is that the particle is removed
from site 1 with rate $\alpha$ and the particle is added to site $L-1$
with rate $\beta$. The flux in such a  system was found in
Ref.~\cite{DEHP}; this gives us
\begin{equation}
\label{VLab}
V_L(\alpha, \beta)=\frac{C_{L-2}(\alpha, \beta)}{C_{L-1}(\alpha,
\beta)}
\end{equation}
where we used the shorthand notation
\begin{equation*}
C_N(\alpha, \beta)=\sum_{p=1}^N \frac{p(2N-1-p)!}{N! (N-p)!}\,\,
\frac{\alpha^{-p-1}-\beta^{-p-1}}{\alpha^{-1}-\beta^{-1}}
\end{equation*}
Plugging $C_0=1$ and $C_1=\alpha^{-1}+\beta^{-1}$ into \eqref{VLab} we
recover  the expression \eqref{V-lame} for the velocity of the bipedal
lame spider; the velocity of the lame tripod is
\begin{equation*}
V_3(\alpha, \beta)=\frac{\alpha^{-1}+\beta^{-1}}
{\alpha^{-2}+\alpha^{-1}\beta^{-1}+\beta^{-2}+\alpha^{-1}+\beta^{-1}}
\end{equation*}
The speed of the infinite-leg spider exhibits an amusing dependence on
the rates $\alpha$ and $\beta$:
\begin{equation}
\label{V-inf}
V_\infty = \left\{
\begin{array}{cl} \displaystyle
1/4 & \mbox{~~for~~} \alpha\geq 1/2,\,\beta\geq 1/2\\  \displaystyle
\alpha(1-\alpha) &  \mbox{~~for~~} \alpha\leq\beta,~~ \alpha<1/2\\
\displaystyle \beta(1-\beta) &  \mbox{~~for~~}
\beta\leq\alpha,~~\beta<1/2
\end{array}
\right.
\end{equation}
Thus if both rates exceed 1/2, the speed attains a universal
(independent of  the rates) maximal value $V_\infty=1/4$. On the other
hand, if at least one of the two extreme legs has the intrinsic speed
less than 1/2, the speed of the entire spider is solely determined by
the slowest leg.

A general explicit expression for the diffusion coefficient is
unknown. There are two special  cases, however, in which the
diffusion coefficient was explicitly calculated \cite{DEM}. One is the
homogeneous spider  ($\alpha=\beta=1$) when $D(L)$ is given by
Eq.~\eqref{DL}; another particular  case corresponds to
$\alpha+\beta=1$ when the diffusion coefficient is
\begin{equation}
\label{DLab}
D_L=\frac{1}{2}\,V_\infty\left\{1-\sum_{k=0}^{L-2}\frac{2 (2k)!}{k!
(k+1)!}\, V_\infty^{k+1}\right\}
\end{equation}
with $V_\infty$ given by Eq.~\eqref{V-inf}; since \eqref{DLab} is
valid on the line  $\alpha+\beta=1$, we have
$V_\infty=\alpha(1-\alpha)=\beta(1-\beta)$. As a consistency check one
can verify that equations \eqref{DLab} and \eqref{D-lame} do agree~: 
setting $L=2$ in the former and $\alpha+\beta=1$ in the latter we
indeed obtain the same result.

The behavior of $D_L$ for the spider with many legs is again
amusing. For the infinite-leg spider, Eq.~\eqref{DLab} yields
\begin{equation}
\label{D-inf-ab}
D_\infty=\frac{1}{2}\alpha\beta|\alpha-\beta| \quad{\rm when}\quad
\alpha+\beta=1
\end{equation}
Thus on the line $\alpha+\beta=1,$ the diffusion coefficient vanishes
only when $\alpha=\beta=1/2$.

The behavior of the diffusion coefficient for the infinite-leg spider
is particularly neat, and it had actually been understood (in the
context of the TASEP) for arbitrary $\alpha$ and  $\beta$. Derrida,
Evans and Mallick \cite{DEM}  found  that
\begin{equation}
\label{D-inf}
\frac{D_\infty}{V_\infty} = \left\{
\begin{array}{cl} \displaystyle
0 & \mbox{~~for~~} \alpha\geq 1/2,\,\beta\geq 1/2\\  \displaystyle
(1-2\alpha)/2 &  \mbox{~~for~~} \alpha<\beta,~~ \alpha<1/2\\
\displaystyle (1-2\beta)/2  &  \mbox{~~for~~}
\beta<\alpha,~~\beta<1/2\\  \displaystyle (1-2\beta)/3  &
\mbox{~~for~~} \alpha=\beta<1/2
\end{array}
\right.
\end{equation}
The discontinuity on the symmetry line $\alpha=\beta<1/2$ is
especially striking.

\section{Quick spiders}
\label{long}

In the previous sections we have considered the simplest possible gaits when the spider's legs can step only to the neighboring sites. In this section we briefly explore the behavior of quick spiders. These spiders (introduced in Ref.~\cite{OS}) differ from previously discussed spiders, namely the legs of a quick spider can jump over several lattice sites at once. The only requirement is to stay within distance $S$ from the other legs. Hence quick spiders can be in the same states as the corresponding global spiders, but more transitions are possible between the states of the quick ones.

The simplest quick spider has two legs always next to each other ($L=2, S=1$). Although such a global spider cannot move, a quick spider can put one leg ahead of the other and can walk this way. Its motion is completely equivalent to a simple random walk, hence its diffusion coefficient is $D=1$. This is generally true for quick spiders with $L$ legs and maximal distance $S=L-1$
\begin{equation}
 D(L, S=L-1) = 1 
\end{equation}

We also computed the diffusion coefficient of bipedal quick spiders with arbitrary $S$. We found
\begin{equation}
\label{DS-jump} 
D(2,S)=\frac{S(S+1)(2S+1)}{6}
\end{equation}
This expression can be derived using the general formula \eqref{D-Tibor}.
For the bipedal spider we can label various configuration by the distance 
$1\leq \ell\leq S$
between the legs
\begin{equation}
\label{conf}
\ldots\circ\bullet\underbrace{\circ\cdots\circ}_{\ell-1}\bullet\circ\ldots
\end{equation}
Take the left leg. It can jump to the left up to distance $S-\ell$; the corresponding 
displacements of the center of mass are $\Delta x=-i/2$ with $1\leq i\leq S-\ell$. The left leg can also jump to the right. The displacements are $\Delta x=i/2$ with $1\leq i\leq \ell-1$, and once it overtakes the right leg, $\Delta x=(\ell+i)/2$ with $1\leq i\leq S$. Taking also into account that all weights are equal, $w_\ell=1/S$, and recalling that jumping of the right leg will give the same contribution, we recast 
\eqref{D-Tibor} into 
\begin{equation}
\label{DiffS-jump} 
D=\frac{1}{4S}\sum_{\ell=1}^{S} \left[\sum_{i=1}^{S-\ell} i^2
+\sum_{i=1}^{\ell-1} i^2+\sum_{i=1}^{S} (\ell+i)^2\right]
\end{equation}
Computing the sum yields the announced result \eqref{DS-jump}.

\section{Interacting Spiders}
\label{inter}

In experiments \cite{exp}, thousands of spiders are released, yet their
density is usually small. Naively, one can anticipate that spiders are
essentially non-interacting. This is correct in the earlier stage,
$t<t_*$, but eventually spiders ``realize'' the presence of other
spiders, and their behavior undergoes a drastic change from diffusive
to a sub-diffusive one. This intermediate stage proceeds up to time
$t^*$ when spiders explore the entire system and then the diffusive
behavior is restored, albeit with a {\em smaller} diffusion
coefficient $\mathcal{D}$. Here we compute $\mathcal{D}$ and estimate
the crossover times $t_*$ and $t^*$.

Let $N$ spiders be  placed on the ring of size $S$. We
assume that the spider density $n=N/S$ is low, $n\ll 1$;
equivalently the typical distance ($n^{-1}$ lattice spacings) between
neighboring spiders is large.

Imagine that we know the diffusion coefficient $D$ of an individual
spider (e.g., for the bipedal spider with $s=3$ we found $D=1/3$ when
each leg hops symmetrically with rates equal to one). Each spider
covers around $\sqrt{D t}$ lattice sites, and equating  $\sqrt{D
t_*}=n^{-1}$ we arrive at the estimate of the lower crossover time
\begin{equation}
\label{time-small}
t_*=\frac{1}{Dn^2}
\end{equation}

The behavior is sub-diffusive in the intermediate time range, $t_*\ll
t\ll t^*$. It is characterized by the $(\lambda t)^{1/4}$ growth of
the covered line \cite{file}; this so-called single-file diffusion has
numerous applications \cite{single,single-sci}. The amplitude
$\lambda$ is found by matching $(D t_*)^{1/2}=(\lambda t_*)^{1/4}$
which in conjunction with \eqref{time-small} yield $\lambda=D/n^2$.

The final behavior is again diffusive. In the long time regime, $t\gg
t^*$, we may interpret each spider as an effective particle hopping to
the right or left with rates $D$. The interaction between spiders is
essentially equivalent to exclusion interaction between particles, and
hence the system reduces to the SEP. We can therefore use
\eqref{D-global} where we should replace $L$ by $N$, and we
must also multiply the result by $D$ since spiders effectively hop
with rates $D$ rather than one. The term in the brackets in
Eq.~\eqref{D-global} reduces to $1-n$; we can replace it by one since
$n\ll 1$. Therefore Eq.~\eqref{D-global} becomes
\begin{equation}
\label{D-many}
\mathcal{D}=N^{-1}D
\end{equation}
Thus exclusion interaction greatly reduces the diffusion
coefficient. This strong cooperative effect emerges even when the
density is arbitrarily small, the only requirement is that there are
many spiders, $N\gg 1$.

The upper crossover time $t^*$ is found by equating  $(\mathcal{D}
t^*)^{1/2}=(\lambda t^*)^{1/4}$. We arrive at
\begin{equation}
\label{time-large}
t^*=\frac{S^2}{D}=N^2 t_*
\end{equation}

Thus the analogy with SEP essentially solves the problem in the
practically important limit when the spider concentration is
low. Neither memory nor the gait play any role, one must merely use
the diffusion coefficient $D$ corresponding to the actual gait  and
computed under the  assumption that the lattice sites are in the product
state.   One should remember, of course, that the SEP regime is
achieved when $t>t^*$; at much earlier times $t>t_*$,  the spiders
mostly hop on the product, and therefore the assumption of full
attachment can become problematic.

\section{Discussion}
\label{disc}

A single spider is a self-interacting object. There are two sources of interaction between the legs:  (i) exclusion (no more than one leg per site), and (ii) legs cannot be too far apart. Is it possible to represent a spider as an effective single particle? The answer is yes --- at least in simple situations, one can treat a spider as a diffusing particle. It is far from trivial, of course, to compute the diffusion coefficient of this particle. Fortunately, natural models of spiders are related to simple exclusion processes. In the course of this work we had an advantage of utilizing some beautiful results and powerful techniques developed in the studies of simple exclusion processes. 

Our models certainly do not take into account all the details of an experimental situation \cite{exp}. For instance, we assumed that the re-attachment of a leg is very quick, so the process is controlled by detachment. Hence spiders remain fully
attached and never leave the surface.  This assumption is important as
our analysis has relied on the permanent presence of spiders on the
surface. Relaxing this assumption does not make the problem
intractable --- indeed in recent analyzes of molecular motors the
complete detachment (unbinding) from cytoskeletal filaments is
allowed, see e.g. \cite{det-frey,det-evans,filament,traffic,mauro}. Further, our
analysis of the many-spider situation in Sec.~\ref{inter} treats the
low density case; the analogy with SEP allowed us to handle the
problem but the assumed permanent presence of the spiders is
particularly questionable in this case.

Perhaps the most serious limitation of our analysis is the disregard of memory --- in experimental realizations \cite{exp} spiders often affect the environment which in turn affect their motion. The non-Markovian nature of this problem calls for a set of new techniques even in the case of a single spider. In one dimension, the influence of memory can be probed analytically for a single bipedal spider \cite{memospiders}, and the replacement of a self-interacting spider by an effective particle remains valid, though this effective particle becomes an excited random walk which distinguishes visited and unvisited sites.

Finally we note that the SEP and its generalizations occur in various biological problems ranging from motion of molecular motors \cite{det-frey,det-evans,filament,traffic,mauro} to protein synthesis \cite{RNA,shaw,chou,chow}. Some models of protein synthesis resemble complicated models of spiders. Another intriguing connection is between spiders and cooperative cargo transport by several molecular motors \cite{cargo}.

\acknowledgments{We are thankful to M. Olah, S. Rudchenko, G. M. Sch\"utz, D. Stefanovic, and M. Stojanovic  for very useful conversations. We also acknowledge financial support to the Program for Evolutionary Dynamics at Harvard University by Jeffrey
Epstein (TA), NIH grant R01GM078986 (TA), and NSF grant CHE0532969 (PLK).}

\appendix

\section{Master equation and fluctuations}
\label{AppA}

In this Appendix, we explain the general formalism, inspired by Ref.~\cite{derrida},  that
allows one to calculate velocities and diffusion contants, and we use  this method  to derive  equation~\eqref{Ds-lame}.

A  spider can be viewed as a homogeneous  Markov process with a finite number of  internal states. The dynamics of the spider  is  encoded in a  Markov Matrix $M$,
where the   non-diagonal matrix element  $ M({C},{C}') $
 represents the rate of evolution from a  configuration 
 ${C}'$  to a different  configuration  ${C}$.
 The quantity    $-M({C},{C}) $ is the  exit-rate
 from configuration  ${C}$. 
  The master equation
 for   $P_t({C})$, 
  the probability of  being in configuration ${C}$ at time $t$, 
  is  then given by 
\begin{equation}
    \frac{d}{dt} P_t({C})  = \sum_{{C}'}
      M({C},{C}') P_t({C}') \, . 
 \label{eq:Markov}
\end{equation}
We now define   $Y_t$  as the  absolute position of the spider's left leg, knowing that at  time $t =0$,   $Y_t = 0$.  Between  $t$ and $t+dt$,  $Y_t$ varies by  the  discrete
amount  $+1, 0$ or $-1$   that depends on the configuration  ${C}'$  at $t$ and on the configuration ${C}$  at $t+dt$.
  The Markov Matrix $M$ can  then be decomposed in three parts corresponding
 to the three possible evolutions of  $Y_t$~:  
 \begin{equation}
M({C},{C}') = 
 M_0({C},{C}') +   M_1({C},{C}')
+   M_{-1} ({C},{C}') \,. 
\label{Markovdecomp}
\end{equation}
For example,  $M_1({C},{C}')$  represents the  transition rate  from  a configuration ${C}'$ to ${C}$
 with the left leg moving  one step
 forward  (this matrix element vanishes otherwise);  $ M_{-1} $ corresponds
 to transitions for  which the left leg moves one step backwards;
 $ M_0 $ encodes transitions in  which the left leg  stays still. 
  We call  $P_t({C}, Y)$ the joint probability
 of being at time $t$ in the configuration ${C}$ and
 having $Y_t = Y$. A master equation,
 analogous to equation~(\ref{eq:Markov}), 
 can  be written for  $P_t({C}, Y)$ as follows~:
\begin{eqnarray}
    \frac{d}{dt} P_t({C}, Y)  &=& \sum_{{C}'} \Big(
   M_0({C},{C}')  P_t({C}', Y)  \nonumber \\
&+&   M_1({C},{C}')  P_t({C}', Y -1) \nonumber \\
&+&   M_{-1} ({C},{C}')  P_t({C}', Y +1)  
        \Big)   \, . 
 \label{eq:Markov2}
\end{eqnarray}
 In terms of   the generating function  $F_t({C})$ defined as
 \begin{equation}
  F_t({C}) =  \sum_{ Y =-\infty}^\infty 
  {\rm e}^{\lambda Y} P_t({C}, Y) \, ,
 \label{eq:defF}
\end{equation}
 the  master equation~(\ref{eq:Markov2}) takes the simpler
 form~:
\begin{equation}
    \frac{d}{dt} F_t({C})  =  \sum_{{C}'} 
  M(\lambda; {C},{C}')   F_t({C}')
   \, , 
 \label{eq:Markov3}
\end{equation}
 where   $ M(\lambda; {C},{C}')$, which  governs the evolution of  $F_t({C})$, 
   is   given by
 \begin{equation}
  M(\lambda) =   M_0  +  {\rm e}^\lambda  M_1  + 
 {\rm e}^{-\lambda}  M_{-1}  \, . 
 \label{eq:defMgamma}
\end{equation}
 We emphasize  that  $M(\lambda)$, 
  is not a Markov matrix for $\lambda \neq 0$
 (the sum of the elements in  a  given column does not vanish).

 In the long time limit, $ t \to \infty$, 
  the behaviour of   $F_t({C})$
 is dominated by the largest eigenvalue $\mu(\lambda)$ 
  of the matrix  $M(\lambda)$. We thus have,  when  $ t \to \infty$,
 \begin{equation}
    \langle  \,  {\rm e}^{\lambda Y_t}\,  \rangle = 
  \sum_{{C}}   F_t({C})  \sim  {\rm e}^{\mu(\lambda)t} 
   \, .
 \label{eq:limF2}
\end{equation}
 This result can be restated more precisely as follows~:
  \begin{equation}
   \lim_{t \to \infty} \frac{1}{t}
 \log   \langle  \,  {\rm e}^{\lambda Y_t}\,  \rangle =  \mu(\lambda) 
   \, .
 \label{eq:limF3}
\end{equation}
The  function $\mu(\lambda)$  contains the complete information about the  
cumulants of  $Y_t$ in the long time limit. For example, the velocity  $V$
and the  diffusion coefficient   $D$ of the spider are given by 
\begin{eqnarray}
 V  &=&  \lim_{t \to \infty} \frac{ \langle Y_t \rangle}{t}  = 
 \frac{ {\rm d}  \mu(\lambda)}{ {\rm d} \lambda} \Big|_{\lambda =0}  
    =  \mu'(0) \, , \label{eq:defV}   \\ 
 D  &=& \lim_{t \to \infty} 
\frac{ \langle Y_t^2 \rangle - \langle Y_t \rangle^2 }{2t}   
  = \frac{ \mu''(0)}{2} \, . \label{eq:defDelta}
 \end{eqnarray}
One therefore needs to calculate the  function  $\mu(\lambda)$. For simple problems such as the bipedal spider with $s=2$,  $\mu(\lambda)$ can be determined explicitely (because $M(\lambda)$ is a 2 by 2 matrix). In general, the  most efficient  technique
 is to perform a perturbative calculation  of  $\mu(\lambda)$ in the vicinity
 of $\lambda =0$ (recall  that $\mu(\lambda)$ vanishes at $\lambda = 0$).  This perturbative approach is very similar to the one used in Quantum Mechanics, the major  difference being that  $M(\lambda)$ which plays the role of the Hamiltonian  is   not, in general, a  symmetric   matrix and  its right eigenvectors are different from its left eigenvectors. By definition, we have
\begin{eqnarray}
  M(\lambda) \,  | \,  \mu(\lambda) \,  \rangle  &=&
  \mu(\lambda) \,  | \,  \mu(\lambda) \,  \rangle  \nonumber  \,  , \\ 
 \langle  \,  \mu(\lambda) \, |   M(\lambda)  &=& 
  \mu(\lambda)   \langle  \,  \mu(\lambda) \, | \, .
 \label{defmulambda}
\end{eqnarray}
Using equations \eqref{eq:defMgamma}, \eqref{eq:defV}, and \eqref{eq:defDelta},
we can write the following perturbative expansions in the vicinity
 of $\lambda = 0$, 
\begin{eqnarray*}
 M(\lambda) &=&    M +  \lambda ( M_1 - M_{-1} ) +
  \frac{\lambda^2( M_1 +  M_{-1} ) }{2}  \ldots         \, \\
     \mu(\lambda)     &=&   V  \lambda   + D {\lambda^2}  + \ldots      \, \\ 
 | \,  \mu(\lambda) \,  \rangle \,  &=& 
  | \, 0 \,  \rangle  +  \lambda  \,  | \, 1 \,  \rangle 
 +    {\lambda^2}  \,  | \, 2 \,  \rangle  + \ldots       \, ,  \\
     \langle  \,  \mu(\lambda) \, |   &=&   \langle  \,  0  \, |  
 + \lambda   \langle  \,  1  \, |   +    {\lambda^2} 
  \langle  \,  2  \, |   +  \ldots   
\end{eqnarray*} 
where $M$ is the original Markov matrix of the system, 
 $  | \, 0 \,  \rangle$ is the  stationary state
 and  $  \langle  \,  0  \, | = (1,1,\ldots,1)$
 is the left ground state of $M$. 
 We now  substitute these perturbative expansions
  in~(\ref{defmulambda}) and identify   the terms with
 the same power of $\lambda$. Using  the left eigenvector
 $ \langle  \,  \mu(\lambda) \, |,$  we obtain
\begin{eqnarray}
\langle  \,  0  \, |  M  &=&   0 \, ,    \label{ordre0}  \\
 \langle  \,  1  \, | M  &=& V \,  \langle  \,  0  \, | 
   - \langle  \,  0  \, | ( M_1 - M_{-1} )  \, ,  \label{ordre1}  \\
   \langle  \,  2  \, | M  &=&   D \, \langle  \,  0  \,  | 
 - \frac{1}{2} \langle  \,  0  \, | ( M_1 + M_{-1} ) \nonumber \\
&\,& +  \, V  \langle  \,  1  \, | 
  -   \langle  \,  1  \, |  ( M_1 - M_{-1} )  \, .  \label{ordre2}
\end{eqnarray} 
 Multiplying  these equations  by the   
 right ground state   $| \, 0 \,  \rangle$  of  $M$,
 and using the fact  that  $ M | \, 0 \,  \rangle = 0 $  and
  $\langle\, 0 \,| \, 0 \,  \rangle  = 1$, the following 
 formulae  for $V$ and $D$ are derived as solvability conditions
 for  Eqs.~\eqref{ordre0}--\eqref{ordre2}~:
\begin{eqnarray}
   V &=& \langle\, 0 \,| M_1 - M_{-1}   | \, 0 \,  \rangle   \, , 
 \label{eqV}\\
  D  &=& \langle\, 1 \,| M_1 -  M_{-1}   | \, 0 \,  \rangle \nonumber\\
 &+& \frac{1}{2}\,\langle\, 0 \,| M_1 +  M_{-1}   | \, 0 \,  \rangle
  -  V   \langle\, 1 \,| \, 0 \,  \rangle \, . 
 \label{eqDelta}
\end{eqnarray} 
We observe that in order to calculate $V$ we only need to know the ground state of $M$. However, the calculation of $D$ requires the knowledge of $\langle\, 1 \,|,$ obtained by solving  the linear equation~(\ref{ordre1}). We remark that similar expressions  can be obtained starting   from the  expansion of  right eigenvector  
$| \,  \mu(\lambda) \,  \rangle$. 

We now specialize this framework  to the case of the heterogeneous bipedal lame spider with $s$ internal states. The Markov Matrix  is then an $s\times s$  matrix
 $M = M_0 + M_1$ since $M_{-1}$ vanishes identically. The matrix $M_0$ is given by 
\begin{equation*}
 M_{0} =  \begin{pmatrix} -\alpha  &  && &0 &\\
\alpha  & -(\alpha + \beta) &  & &\\
& \,\,  \beta &  -(\alpha + \beta)  &  &\\
& &   \ddots &  \ddots & &\\
& &  & \beta  &    -(\alpha + \beta) & \\
& 0 & & & \beta & -\alpha \end{pmatrix}
\end{equation*} 
and the matrix $M_1$ is  
\begin{equation*}
 M_{1} =  \alpha  \begin{pmatrix} 0  & 1 & & & 0 &\\
 & 0 & 1 &   &\\
&  & 0 & 1 &  & \\
& & &\ddots & \ddots \\ 
&  & &  & 0 & 1  \\
& 0 & &  &  &  0 \end{pmatrix}
\end{equation*}
The stationary state of  $M$ is 
$  | \, 0 \,  \rangle = (p_0, p_1,\ldots,p_{s-1})$ with 
\begin{equation}
 p_k = \frac{\alpha -\beta}{\alpha^s -\beta^s}\, 
 \alpha^{s-k-1}\beta^{k} \,\,  \hbox{ for }\,\,\, k=0,\ldots,s-1 \, .
 \label{eqpk}
\end{equation}
This expression,  together with \eqref{eqV}, 
 leads to the  formula~\eqref{V-lame-gen} for  the spider velocity.

  In order to derive the expression of the
 diffusion  coefficient, we need to solve equation~\eqref{ordre1}.
 One can verify that the solution of this equation is given
 by  $   \langle  \,  1  \, |   = (q_0, q_1,\ldots,q_{s-1})$ where 
  \begin{equation}
 q_k =  (k+1)\frac{ V -\alpha}{\beta - \alpha}  + 
 \frac{\alpha\beta - \alpha V}{(\beta - \alpha)^2}
  \Big( 1 - \left(\frac{\alpha}{\beta}\right)^k  \Big)
\label{eqqk}
\end{equation}
for  $k=0,\ldots,s-1 \,.$ Inserting equations~(\ref{eqpk}) and~(\ref{eqqk}) into the general  expression~\eqref{eqDelta} leads to the formula~\eqref{Ds-lame}. 

We also used the above method to determine the velocity and the diffusion coefficient for centipedes with $s=3$. The results (Sect.~\ref{main}) were obtained by explicitly constructing the matrices $M_0, M_1$, and $M_{-1}$, and performing exact computations using {\it Maple}. These computations are feasible when the number of legs is sufficiently small. (The total number of configurations is $3^{L-1}$ for centipedes with $s=3$, and hence the order of matrices $M_0, M_1,M_{-1}$ quickly grows with $L$.)

\section{Generalized detailed balance relation}
\label{AppB}

For the symmetric spider, the three  matrices $M_0$, $M_1$ and  $M_{-1}$,  introduced in \eqref{Markovdecomp} to take into  account  the total displacement  of the spider, satisfy the following detailed balance relation  
\begin{equation}
  M_y({C},{C}') P^{eq}({C}') = 
  M_{-y}({C}',{C}) P^{eq}({C})
  \label{YDetBal}
\end{equation}
where the  equilibrium measure is denoted by $P^{eq}$ and $y =0,\pm 1$.
Equation \eqref{YDetBal} implies that the velocity of the spider vanishes. Consider now a  spider driven out of  equilibrium with  a non-vanishing mean velocity.  Suppose however, that for  the  model  under consideration  there exists a real number 
$\epsilon$ such that the following generalized detailed balance relation is satisfied
\begin{equation}
    M_y(\mathcal{C},\mathcal{C}') P^{eq}(\mathcal{C}') = 
  M_{-y}(\mathcal{C}',\mathcal{C}) P^{eq}(\mathcal{C}) \exp(\epsilon y) \, ,
 \label{gendetbal}
\end{equation}
 Here again,  $ P^{eq} $ is the  equilibrium measure corresponding
 to the symmetric spider.
 From relation~(\ref{gendetbal}) it is a matter of elementary
 algebra to prove that  the spectra  of  $M(\lambda)$
 and   of  $M(-\epsilon  -\lambda)$  are identical. Therefore
 \begin{equation}
 \mu(\lambda) =  \mu(-\epsilon - \lambda) \, .
 \label{eq:GC}
  \end{equation}
This relation, which is a special case of the general {\it Fluctuation
 Theorem} valid for systems far from equilibrium
 \cite{ECM93,ES94,GCohen}, 
  was derived for stochastic systems by 
  Lebowitz and Spohn  \cite{LebSpohn}.
   Close to equilibrium, when  $\epsilon \ll 1$,
   we can expand equation~(\ref{eq:GC}) for small 
  $\lambda$ and  $\epsilon$. We find at lowest order
\begin{equation}
  \mu''(0) =   \frac{ \mu'(0)}{\epsilon}  \,\,\,\, i.e.,   \,\,\,\, 
  D  =  \frac{ V}{\epsilon} \, ,
\label{eq:FDT}
  \end{equation}
which is nothing  but the classical fluctuation-dissipation relation between diffusion and mobility.


\begin{thebibliography}{99}


\bibitem{seeman} H.~Yan, X.~Zhang, Z.~Chen, and N.~C.~Seeman, Nature
  {\bf 415}, 62--65 (2002); W.~B.~Sherman and N.~C.~Seeman,  Nano
  Lett. {\bf 4}, 1203--1207 (2004);  N.~C.~Seeman, Trends
  Biochem. Sci. {\bf 30}, 119 (2005).

\bibitem{pierce} J.-S.~Shin and N.~A.~Pierce, J. Am.\ Chem.\ Soc. {\bf
  126}, 10834 (2004).

\bibitem{DNA_motor}   W.~M.~Shu et al., J. Am.\ Chem.\ Soc. {\bf
  127},  17054 (2005).

\bibitem{review}  E. R. Kay, D. A. Leigh, and F. Zerbetto, Angew.\
   Chem.\ Int.\ Ed. {\bf 46}, 72--191  (2007).

\bibitem{exp} R.~Pei, S.~K.~Taylor, D.~Stefanovic, S.~Rudchenko,
  T.~E.~Mitchell,  and M.~N.~Stojanovic, J. Am.\ Chem.\ Soc. {\bf
  128}, 12693 (2006).

\bibitem{markov} For Markov processes, the future is determined by the
   present \cite{van}.

\bibitem{van}  N. G. Van Kampen, {\it Stochastic Processes in Physics
   and Chemistry}  (Elsevier, Amsterdam, 2003).

\bibitem{memospiders}
  T.~Antal and P.~L.~Krapivsky, arXiv:0705.2596.

\bibitem{models} Mathematical models of  animal locomotion are reviewed by
   P.~Holmes, R.~J.~Full, D.~Koditschek, and J.~Guckenheimer,  SIAM
   Rev. {\bf 48}, 207--304 (2006).

\bibitem{OS} 
   M.~J.~Olah and D.~Stefanovic, in preparation. 

\bibitem{motors} J.~Howard, {\it Mechanics of Motor Proteins and the
   Cytoskeleton} (Sinauer Associates, Sunderland, MA, 2001).

\bibitem{FK}    M. E. Fisher and A. B. Kolomeisky, Physica A {\bf
   274}, 241 (1999); A. B. Kolomeisky, J. Chem. Phys. {\bf 115}, 7253
   (2001).

\bibitem{D}    B.~Derrida, J.\ Stat.\ Phys. {\bf 31}, 433--450 (1983).

\bibitem{DDM}    B.~Derrida, E.~Domany, and D.~Mukamel, J.\ Stat.\
   Phys. {\bf 69}, 667 (1992).

\bibitem{gilles}    E.~Duchi and G.~Schaeffer, J.\ Combin.\ Theory A
  {\bf 110}, 1 (2005).

\bibitem{DEM}    B.~Derrida, M.~R.~Evans, and K.~Mallick, J.\ Stat.\
   Phys. {\bf 79}, 833 (1995).

\bibitem{DPramana} B.~Derrida, 
 Pramana-J.~Phys. {\bf 64},  695 (2005).

\bibitem{spohn} H.~Spohn, {\it Large Scale Dynamics of Interacting
   Particles}  (Springer-Verlag, New York, 1991).

\bibitem{derrida} B.~Derrida, Phys.\ Rep. {\bf 301}, 65 (1998).

\bibitem{liggett}  T.~M.~Liggett, {\it Stochastic Interacting Systems:
   Contact, Voter, and Exclusion Processes} (Springer-Verlag, New
   York, 1999).

\bibitem{gunter} G.~M.~Sch\"utz, in: {\it Phase Transitions and
   Critical Phenomena},  eds. C.~Domb and J.~L.~Lebowitz, vol. 19
   (Academic Press, San Diego, 2001).

\bibitem{gunter2}   
        G. Sch\"utz and S. Sandow, Phys.\ Rev.\ E {\bf 49}, 2726 (1994).

\bibitem{DMal}    B.~Derrida  and K. Mallick, J.\ Phys. A
   {\bf 30}, 1031 (1997).

\bibitem{DE}    B.~Derrida, M.~R.~Evans, and D.~Mukamel, J.\ Phys. A
   {\bf 26}, 4911 (1993).

\bibitem{DEHP}    B.~Derrida, M.~R.~Evans, V. Hakim, and V. Pasquier,
   J.\ Phys. A {\bf 26}, 1493 (1993).

\bibitem{file}      D.~G.~Levitt, Phys. Rev. A {\bf 8}, 3050 (1973);
   P. M. Richards, Phys. Rev. B {\bf 16}, 1393 (1977); P. A. Fedders,
   Phys. Rev. B {\bf 17}, 40 (1978); S. Alexander and P. Pincus,
   Phys. Rev. B {\bf 18}, 2011 (1978).

\bibitem{single}  
  B. Alberts et al., {\it Molecular Biology of the Cell} (Garland, New
  York, 1994); N. Y. Chen, T. F. Degnan, and C. M. Smith, {\it
  Molecular Transport and Reaction  in Zeolites} (VCH, New York, 1994).

\bibitem{single-sci}      Q.-H.~Wei,  C.~Bechinger, and P.~Leiderer,
  Science {\bf 287}, 625 (2000).

\bibitem{det-frey}  A. Parmeggiani, T. Franosch, and E. Frey,
  Phys. Rev. Lett. {\bf 90}, 086601 (2003).

\bibitem{det-evans}  M. R. Evans, T. Hanney, and Y. Kafr\'i,
  Phys. Rev. E {\bf 70}, 066124 (2004).

\bibitem{filament}  S.~Klumpp, M.~J.~I.~M\"uller, and R.~Lipowsky,
   in {\it Traffic and Granular Flow '05}, ed. A. Schadschneider et al. 
   (Springer, Berlin, 2007), pp. 251--261.

\bibitem{traffic}  P.~Greulich, A.~Garai, K.~Nishinari,
   A.~Schadschneider, and D.~Chowdhury, physics/0612054.

\bibitem{mauro}  M.~Mobilia, T.~Reichenbach, H.~Hinsch, T.~Franosch,
   and E.~Frey,  cond-mat/0612516.

\bibitem{RNA}  A. C. Pipkin and J. H. Gibbs, Biopolymers {\bf 4}, 3
    (1966); C. T. MacDonald,  J. H. Gibbs, and A. C. Pipkin,
    Biopolymers {\bf 6}, 1 (1968); C. T. MacDonald and J. H. Gibbs,
    Biopolymers {\bf 7}, 707 (1969).
    
\bibitem{shaw}
  L. B. Shaw, R. K. P. Zia, and K. H. Lee,
  Phys.\ Rev.\ E {\bf 68}, 021910 (2003);
  L.~B.~Shaw, A.~B.~Kolomeisky, and K.~H.~Lee, J.\ Phys.\ A {\bf 37}, 2105 (2004);
  L.~B.~Shaw, J.~P.~Sethna, and K.~H.~Lee, Phys.\ Rev.\ E {\bf 70}, 021901 (2004).

\bibitem{chou} T.~Chou and G.~Lakatos,   Phys.\ Rev.\ Lett. {\bf 93}, 198101 (2004).

\bibitem{chow} A.~Basu and D.~Chowdhury,   Phys.\ Rev.\ E {\bf 75}, 021902 (2007).
    
\bibitem{cargo}  S.~Klumpp and R.~Lipowsky, Proc. Natl. Acad. Sci. USA
   {\bf 102}, 17284 (2005); M. Vershinin et al.,
   Proc. Natl. Acad. Sci. USA {\bf 104}, 87 (2007).


 \bibitem{ECM93}
 D.~J.~Evans, E.~G.~D.~Cohen, and G.~P.~Morriss,  
  Phys. Rev. Lett.  {\bf 71}, 2401 (1993).


 \bibitem{ES94}
 D.~J.~Evans and D.~J.~Searles, 
  Phys. Rev. E  {\bf 50}, 1645 (1994).


\bibitem{GCohen}
 G.~Gallavotti and E.~G.~D.~Cohen,
   Phys. Rev. Lett.  {\bf 74}, 2694 (1995); 
 J. Stat. Phys. {\bf 80}, 931 (1995).


\bibitem{LebSpohn} J. L. Lebowitz and H.~Spohn, 
  J. Stat. Phys.  {\bf 95}, 333  (1999).

\end{thebibliography}
\end{document}